\documentclass[twocolumn,prb,showpacs,preprintnumbers,amsmath,amssymb]{revtex4}

\usepackage{graphicx}
\usepackage{dcolumn}
\usepackage{bm}
\usepackage{epsfig,float,afterpage,amssymb,wrapfig,psfrag}

\newcommand{\be}{\begin{equation}}
\newcommand{\cc}{\mbox{\scriptsize{c}}}

\newcommand{\ii}{\mbox{\scriptsize{i}}}

\newcommand{\ee}{\end{equation}}
\newcommand{\bea}{\begin{eqnarray}}

\newcommand{\eea}{\end{eqnarray}}

\newcommand{\ssz}{\scriptsize}
\newcommand{\rmd}{\mbox{d}}
\newcommand{\Or}{\mbox{O}}

\newcommand{\scrG}{{\cal G}}

\newcommand{\w}{\omega}
\newcommand{\K}{\mbox{\scriptsize{K}}}
\newcommand{\m}{\mbox{\scriptsize{m}}}
\newcommand{\wm}{\w_{\m}}
\newcommand{\wk}{\w_{\K}}
\newcommand{\wl}{\w_{\mbox{\scriptsize{L}}}}
\newcommand{\im}{\mbox{i}}
\newcommand{\subi}{\mbox{\scriptsize{I}}}
\newcommand{\subr}{\mbox{\scriptsize{R}}}
\newcommand{\SUBI}{\mbox{\scriptsize{I}}}
\newcommand{\SUBR}{\mbox{\scriptsize{R}}}

\newcommand{\up}{\uparrow}

\newcommand{\st}{\tilde{\Sigma}}

\newcommand{\Tt}{\tilde{T}}
\newcommand{\wt}{\tilde{\w}}
\newcommand{\down}{\downarrow}

\DeclareMathAlphabet{\bi}{OML}{cmm}{b}{it}
\newcommand{\kk}{\bi{k}}

\newcommand{\sts}{\tilde{\Sigma}_{\sigma}}

\newcommand{\ra}{\rightarrow}

\renewcommand{\Or}{\cal{O}\rm}
\newcounter{saveeqn}

\newcommand{\PR}{{\it Phys. Rev.}}
\newcommand{\jpcm}{{\it J. Phys.: Condens. Matter}}

\newcommand{\seceq}{\setcounter{equation}{0}}

\usepackage{graphics}
\begin{document}

\title{Local quantum critical point in the pseudogap Anderson model: finite-$T$ dynamics
 and $\w/T$ scaling}
\author{Matthew T Glossop}
\altaffiliation[Current address: ]{Department of Physics, University of Florida,
Gainesville, Florida 32611-8440, USA.} 
\author{Gareth E Jones} 
\author{David E Logan}
\email{dlogan@physchem.ox.ac.uk}

\date{\today}

\affiliation{Oxford University, Physical and Theoretical
Chemistry Laboratory, South Parks Road, Oxford OX1 3QZ, UK.}

\begin{abstract}
The pseudogap Anderson impurity model is a paradigm for
locally critical quantum phase transitions.  Within the framework of the
local moment approach we study its finite-$T$ dynamics, as embodied in the
single-particle spectrum, in the vicinity of the symmetric quantum
critical point (QCP) separating generalized Fermi-liquid (Kondo screened)
and local moment phases.  The scaling spectra in both phases, and at the
QCP itself, are obtained analytically.  A key result is that pure
$\w/T$-scaling obtains at the QCP, where the Kondo resonance has just
collapsed.  The connection between the scaling spectra in either phase and
that at the QCP is explored in detail.
\end{abstract}


\maketitle

\section{Introduction}

Proximity to a magnetic quantum critical point is one mechanism
for non-Fermi liquid behaviour in heavy fermion metals, a topic of
great current interest \cite{gs,coleman}. Experimentally, the 
temperature below which certain heavy fermion materials become
antiferromagnetically ordered can be tuned to $T = 0$ either chemically 
by doping (e.g. CeCu$_{6-x}$M$_x$ with M =
Au or Ag \cite{vl,heuser,kuch1}) or through applied pressure (e.g. CeIn$_3$
\cite{ndm}) or a magnetic
field
(e.g. YbRh$_2$Si$_2$ \cite{ot}).  The transition point then becomes a
quantum 
critical
point and leads to non-Fermi liquid behaviour extending over
large regions of the phase diagram.

An example that has received much attention is  CeCu$_{6-x}$Au$_x$
\cite{vl,schroder,os}.  For $x
< x_{\cc}\simeq 0.1$ it is a paramagnet with properties consistent with a
highly renormalized Landau Fermi liquid.  The localized magnetic moments
provided by the Ce atoms, manifest as a Curie-Weiss law at high
temperatures, are screened by conduction electrons at low temperatures in
a lattice analogue of the familiar Kondo effect for magnetic impurities in
metals.  For $x>x_{\cc}$ by contrast, coupling between local moments wins out and the
ground state has antiferromagnetic long-ranged order.

The observed non-Fermi liquid behaviour of  CeCu$_{5.9}$Au$_{0.1}$
\cite{schroder,os}, probed by neutron scattering experiments, cannot however be
understood in terms of the standard Hertz-Millis mean-field description of a magnetic
quantum phase transition based on long wavelength fluctuations of the
order parameter 
\cite{hertz,millis,moriya} --- the spin dynamics are found to display fractional exponents throughout the Brillouin zone, and satisfy pure
$\w/T$-scaling.  This suggests the emergence at the QCP of local
moments that are temporally critical, and that temperature is the only
energy scale in the problem, cutting of critical quantum fluctuations
after a correlation time on the order of $\hbar/k_{\mbox{\ssz{B}}}T$.

An alternative picture, where critical {\it local} fluctuations coexist
with spatially extended ones, has recently been proposed to explain the
above results.  Si {\it et al} \cite{sinat} have shown that such a local
quantum
critical
point arises in the Kondo lattice model through competition between the
Kondo exchange coupling --- describing the interaction of conduction
electron spin and local moment at each lattice site --- and the RKKY
interaction, which characterizes the coupling between local moments at
different lattice sites.  At this local QCP the lattice system reaches its
magnetic ordering transition at precisely the same point that the Kondo
screening becomes critical.
Thus the Kondo lattice model, and the self-consistent Bose-Fermi Kondo
model to which it reduces with an extended dynamical mean-field theory
treatment, are the focus of much current theoretical research
\cite{sinat,smith,zhu,grempel,zhuetal,burdin}.

However, significant progress in understanding local QCPs
has followed from the study of a simpler problem: the pseudogap Anderson
[Kondo] impurity model (PAIM), which serves as a paradigm for locally
critical quantum phase transitions, and is the focus of this work.
Proposed over a decade ago by Withoff and Fradkin \cite{wf}, the model
represents a
generalization of the familiar Anderson [Kondo] impurity model
(AIM) \cite{pwa}
describing a single magnetic impurity embedded in a non-interacting
metallic host (for reviews see Ref.\ \onlinecite{hew,gehring}), and itself enjoying a
resurgence of interest in the context of quantum dots and STM
experiments \cite{revival,qdots}.

In the PAIM the fermionic host is described by a power-law density of
states vanishing at the Fermi level: $\rho(\w) \propto |\w|^r$ with $r>0$ 
($r=0$ recovers the case of a metallic host).
Many specific examples of pseudogap systems \emph{per se} have also been   
proposed, including various zero-gap semiconductors \cite{vp}, one-dimensional
interacting systems \cite{voit}, and the Zn- and Li-doped cuprates
\cite{voj,vb02}.  And the possibility
that pseudogap physics might be realized in a quantum dot system has also
recently been suggested \cite{hopkinson}.
The essential physics of the PAIM is by now well understood via a number of
techniques, e.g.\ perturbative scaling
\cite{wf,ki96,gbi96}, numerical renormalization group
\cite{cj,bph97,gbi98,bglp,ingsi02,vb02}, perturbative renormalization 
group \cite{mvrecent}, local moment approach
\cite{bglp,lg00,gl03,gl032}, large-$N$ methods \cite{wf,cf,ingsi98,voj},
non-crossing approximation \cite{bork} and bare perturbation theory in the
interaction strength $U$ \cite{gl00}. 
In contrast to the regular metallic AIM, which exhibits Fermi liquid physics and a
Kondo effect for all $U$ [or $J$ in its Kondo model limit], the 
PAIM displays in general critical
local moment fluctuations and a destruction of the Kondo effect at a
finite critical $U_{\cc}$ [or $J_{\cc}$].  The quantum critical point,
with spin correlations that are critical in time yet spatially local,
separates a Fermi liquid Kondo-screened phase ($U<U_{\cc}$) from a local
moment phase with the characteristics of a free spin-$1/2$.  The key
features of the model are summarized in \S 2.
\newline\indent In an effort to understand better the physics of critical local moment
fluctuations, the most recent work on pseudogap impurity models has
naturally focused on the QCP itself \cite{voj,ingsi02,gl03,mvrecent},
which has
been established as
a non-Fermi liquid interacting fixed point for $r<1$ (the upper critical
dimension of the problem).  For example, Ingersent and Si
\cite{ingsi02} have recently
shown that for $r < 1$ the dynamical spin susceptibility at the critical
point exhibits $\w/T$-scaling with a fractional exponent, which features
closely parallel those of the local QCP of the Kondo lattice.  A very
recent perturbative renormalization group study \cite{mvrecent} has
developed critical
theories for the transition, confirming that, for $0<r<1$ the transition
is described by an interacting field theory with universal local moment
fluctuations and hyperscaling, including $\w/T$-scaling in the dynamics.
\newline\indent In this paper, using the local moment approach (LMA), we develop an analytical description of finite-$T$  dynamics in the vicinity of the QCP, as embodied in the local single-particle 
spectrum $D(\w)$.
Developed originally to describe metallic AIMs at $T = 0$ \cite{let,dl01,gl02},
and
subsequently extended to handle finite temperatures \cite{ld02} and magnetic
fields
\cite{ld01a,ldb01}, the LMA has also been applied to the PAIM \cite{lg00,gl03,gl032}
leading to a number of
new predictions that have since been confirmed by NRG calculations
\cite{bglp}.  We also add that the approach can encompass lattice-based  
models
within dynamical mean-field theory, e.g. the periodic Anderson model
appropriate to the paramagnetic metallic phase of heavy fermion
materials \cite{raja,ves,raja2}.
\newline\indent The paper is organized as follows.  After the necessary background,
section 2, we present in section 3 a number of exact results for the
scaling spectra in both phases of the model, and at the QCP itself, using
general scaling arguments.  In section 4 a brief description is given of
the finite-$T$ local moment approach to the problem, along with a
numerical demonstration of scaling of the single-particle dynamics.  An
analytical description, arising within the LMA, of the finite-$T$
scaling spectrum for both phases and at the QCP is given in section 5, and
is found to be in excellent agreement with the numerics. The connection
between the scaling spectra in either phase and that at the QCP itself is
explored fully,  $\w/T$-scaling of $D(\w)$ at the QCP being one key result.  
The paper concludes with a brief summary.

\seceq
\section{Background}
\label{background} The Hamiltonian for an AIM is given in standard
notation by \be
\hat{H}=\sum_{\kk,\sigma}\epsilon_{\kk}\hat{n}_{\kk\sigma}+\sum_{\sigma}(\epsilon_{\ii}+\frac{U}{2}\hat{n}_{\ii-\sigma})\hat{n}_{\ii\sigma}
+\sum_{\kk,
\sigma}V_{\ii\kk}(c^{\dagger}_{\ii\sigma}c_{\kk\sigma}+\mbox{h.c.}) \ee
where the first term describes the non-interacting host with
dispersion $\epsilon_{\kk}$ and density of states  $\rho(\omega) =
\sum_{\kk}\delta(\omega - \epsilon_{\kk})$ (with $\omega =0$ the
Fermi level). The second term describes the correlated impurity,
with energy $\epsilon_{\ii}$ and local interaction $U$, and the
third the host-impurity coupling (via \rm $V_{\ii\kk} \equiv V$).

  For a conventional metallic host the Fermi level density of states $\rho(0)
\neq 0$, and low-energy states are always available for screening. In
consequence the impurity spin is quenched and the system a Fermi liquid for
all $U$ (see e.g.\ \cite{hew}). For the pseudogap AIM (PAIM) by contrast the
host density of states is soft at the Fermi level, $\rho(\omega) \propto
|\omega|^{r}$ ($r>0$) \cite{wf}. Due to the depletion of host states around
the Fermi level, the PAIM exhibits a quantum phase transition
\cite{wf,ki96,gbi96,cf,ingsi98,voj,cj,bph97,gbi98,bglp,vb02,ingsi02,lg00,gl03,gl032}
at a critical $U=U_{\cc}(r)$, the quantum critical point (QCP) separating a
degenerate local moment (LM) phase for $U>U_{\cc}$ from a `strong coupling'
or generalized Fermi liquid (GFL) phase, $U < U_{\cc}$, in which the impurity
spin is locally quenched and a Kondo effect manifest. As $r \rightarrow 0$
the critical $U_{\cc}(r)$ diverges ($\sim 1/r$)
\cite{bph97,gbi98,bglp,lg00,gl03,gl032}, symptomatic of the absence of a
transition for the conventional metallic AIM corresponding to $r=0$, and
whose behaviour is recovered smoothly in the $r \rightarrow 0$ limit
\cite{lg00,gl03,gl032}. In the present paper we focus on the so-called
symmetric QCP that arises in both the particle-hole (p-h) symmetric and
asymmetric PAIM \cite{gbi98}. To that end we consider explicitly the p-h
symmetric model, with $\epsilon_{\ii} = -\frac{1}{2}U$ (and impurity charge
$n_{\ii} = \sum_{\sigma}\langle\hat{n}_{\ii\sigma}\rangle = 1$). In this case the QCP is
known to separate GFL/LM phases for all $0<r<\frac{1}{2}$
\cite{bph97,gbi98,bglp,lg00,gl03} (for $r>\frac{1}{2}$ the LM phase alone
arises for any $U >0$).

  Our focus here is on single-particle dynamics at finite-$T$, embodied
in the (retarded) impurity Green function $G(\omega,T)$
($\leftrightarrow
-\im\theta(t)\langle\{c_{\ii\sigma}(t),c_{\ii\sigma}^{\dagger}\}\rangle$),
with  $D(\omega,T) = -\frac{1}{\pi}$Im$G(\omega,T)$ the
single-particle spectrum.  $G(\omega,T)$ may be expressed as
\begin{equation}
G(\omega,T) = [\omega^{+} - \Delta(\omega) - \Sigma(\omega,T)]^{-1}
\end{equation}
with $\omega^{+} = \omega + \im 0^{+}$ and $\Sigma(\omega,T) =
\Sigma^{\subr}(\omega,T) - \im\Sigma^{\subi}(\omega,T)$ the
conventional single self-energy (defined to exclude the trivial
Hartree term which precisely cancels $\epsilon_{\ii} =
-\frac{1}{2}U$), such that $\Sigma(\omega,T) =
-[\Sigma(-\omega,T)]^{*}$ by p-h symmetry. All effects of
host-impurity coupling at the one-electron level are embodied in
the hybridization function  $\Delta(\omega) = \sum_{\kk}
V^{2}_{\ii\kk} [\omega^{+} - \epsilon_{\kk}]^{-1}$
$=\Delta_{\SUBR}(\omega) - \im \Delta_{\SUBI}(\omega)$. For the
PAIM considered here, $\Delta_{\SUBI}(\omega) = \pi V^{2}
\rho(\omega)$ is given explicitly by $\Delta_{\SUBI}(\omega) =
\Delta_{0}(|\omega|/\Delta_{0})^{r}\theta(D-|\omega|)$ with
$\Delta_{0}$ the hybridization strength, $D$ the bandwidth and
$\theta(x)$ the unit step function. Throughout the paper we
take $\Delta_{0} \equiv 1$ as the energy unit,  i.e.\
\begin{equation}
\Delta_{\SUBI}(\omega) = |\omega|^{r}\theta(D-|\omega|).
\end{equation}
The real part of the hybridization function follows simply from Hilbert transformation,
\begin{equation}
\Delta_{\SUBR}(\omega) = -
\mbox{sgn}(\omega)\left[\beta(r)\Delta_{\SUBI}(\omega) + {\cal{O}}
\left(\frac{|\omega|}{D}\right)\right]
\end{equation}
with $\beta(r) = \mbox{tan}(\frac{\pi r}{2})$. That this form for
the hybridization is simplified is of course irrelevant to the
low-energy scaling behaviour of the problem, in the same way that
the usual `flat band' caricature of the metallic host is
immaterial to the intrinsic Kondo physics of the metallic AIM
\cite{hew}.

  We now summarize key characteristics of the problem at $T=0$
\cite{wf,ki96,gbi96,cf,ingsi98,voj,cj,bph97,gbi98,bglp,vb02,ingsi02,lg00,gl03,gl032}
that are required in the remainder of
the paper, including the essential manner in which the underlying transition
is directly apparent in single-particle dynamics.

\subsection{GFL phase}
 For \it all \rm $U < U_{\cc}(r)$ in the GFL
phase ($0<r<\frac{1}{2}$), the leading low-$\omega$ behaviour of
$D(\omega,0)$ is entirely unrenormalized from the non-interacting
limit and given by
\begin{equation}
\pi D(\omega,0) \stackrel{|\omega| \rightarrow 0}{\sim}
\mbox{cos}^{2}(\frac{\pi r}{2}) |\omega|^{-r}.
\end{equation}
This is an exact result \cite{gl00}. It arises because
$\Sigma^{\SUBR/\SUBI}(\omega,0)$ vanish as $\omega \rightarrow 0$
more rapidly than the hydridization ($\propto |\omega|^{r}$),
reflecting in physical terms the perturbative continuity to the
non-interacting limit that in essence defines a Fermi liquid. The
leading low-$\omega$ behaviour of $\Sigma^{\SUBR}(\omega,0)$ is in
fact given by
\begin{equation}
\Sigma^{\SUBR}(\omega,0) \stackrel{\omega \rightarrow 0}{\sim}
-\left(\frac{1}{Z} -1\right) \omega
\end{equation}
as one would cursorily expect from p-h symmetry 
($\Sigma^{\SUBR}(\omega,0) = -\Sigma^{\SUBR}(-\omega,0)$); with the
quasiparticle weight (or mass renormalization) $Z$ defined as usual by $Z = [1-
(\partial\Sigma^{\SUBR}(\omega,0)/\partial\omega)_{\omega
=0}]^{-1}$. By virtue of equation (2.5) the most revealing
expos\'e of GFL dynamics at $T=0$ lies in the modified spectral
function ${\cal{F}}(\omega,0)$ \cite{bglp,lg00,gl03,gl032}, defined generally by
\begin{equation}
{\cal{F}}(\omega,T) =\pi\ \mbox{sec}^{2}(\mbox{$\frac{\pi
r}{2}$})|\omega|^{r} D(\omega,T)
\end{equation}
such that ${\cal{F}}(0,0) = 1$ (i.e.\ the $T=0$ modified spectrum
is `pinned' at the Fermi level, a result that reduces to the
trivial dictates of the Friedel sum rule for the $r=0$ metallic
AIM \cite{hew}).

  The Kondo resonance
is directly apparent in ${\cal{F}}(\omega,0)$ \cite{bglp,lg00,gl03,gl032} ---
see e.g.\
figure 16 of Ref. \onlinecite{lg00}, and figure 3 below.
Indeed on visual inspection it is barely
distinguishable from its $r=0$ metallic counterpart, to which it
reduces for $r=0$. The resonance is naturally characterised by a
low-energy Kondo scale $\wk$, usually defined in practice by the
width of the resonance \cite{bglp}. Most importantly, as $U
\rightarrow U_{\cc}-$ and the GFL$\rightarrow$LM transition is
approached, the Kondo scale becomes arbitrarily small (vanishing
at the QCP itself, where the Kondo resonance collapses).  For
$r\ll 1$ it is well-established from previous LMA work and NRG
calculations \cite{bglp,lg00,gl03,gl032} that $\wk$ vanishes according to
$\wk\propto(1-U/U_{\cc}(r))^{\frac{1}{r}}$.

In consequence, in the vicinity of the transition,
${\cal{F}}(\omega,0)$ exhibits \emph{universal scaling behaviour} in
terms of $\omega/\wk$ \cite{bglp,lg00}. It is of course in this
scaling behaviour that the underlying quantum phase transition is
directly manifest \cite{gl03,gl032}.

\subsection{LM phase}
 The local moment phase is more subtle than naive expectation
might suggest. Here it is known, originally from NRG calculations
\cite{bph97}, that the leading low-$\omega$ behaviour of
$D(\omega,0)$ is
\begin{equation}
D(\omega,0) \stackrel{|\omega| \rightarrow 0}{\sim} c|\omega|^{r}
\end{equation}
(with Re$G(\omega,0) \sim
-\mbox{sgn}(\omega)\pi\beta(r)D(\omega,0)$ following directly from
Hilbert transformation). From this alone, simply by inverting
equation (2.2), the leading low-$\omega$ behaviour of the single
self-energy $\Sigma^{\SUBR}(\omega,0)$ is given by
\begin{equation}
\Sigma^{\SUBR}(\omega,0) \stackrel{|\omega| \rightarrow 0}{\sim}
\mbox{sgn}(\omega) \mbox{sin}(\pi r) \frac{1}{2\pi D(\omega,0)}
\hspace{0.3cm} \propto |\omega|^{-r}
\end{equation}
(and likewise $\Sigma^{\SUBI}(\omega,0) \propto
\Sigma^{\SUBR}(\omega,0)$). This divergent low-$\omega$ behaviour
is radically different from the simple analyticity of its
counterpart equation (2.6) in the GFL phase, and illustrates the
basic difficulty:
the need to capture within a common framework such distinct behaviour
in the GFL and LM phases, indicative
of the underlying phase transition and as such requiring an inherently
non-perturbative description. We do not know of any theoretical approach based on
the conventional single self-energy that can handle this issue.

There is however nothing sacrosanct in direct use of the single
self-energy, which is merely defined by the Dyson equation
implicit in equation (2.2). Indeed for the doubly-degenerate LM
phase, a physically far more natural approach would be in terms of
a two-self-energy description (as immediately obvious by
consideration e.g.\ of the atomic limit --- the `extreme' LM
case). Here the rotationally invariant $G(\omega,T)$ is expressed
as
\begin{equation}
G(\omega,T) = \frac{1}{2} [G_{\uparrow}(\omega,T) + G_{\downarrow}(\omega,T)]
\end{equation}
with the $G_{\sigma}(\omega,T)$ given by
\begin{equation}
G_{\sigma}(\omega,T) = [\omega^{+} - \Delta(\omega) -
\tilde{\Sigma}_{\sigma}(\omega,T)]^{-1}
\end{equation}
in terms of spin-dependent self-energies
 $\tilde{\Sigma}_{\sigma}(\omega,T)$ ($=
\tilde{\Sigma}_{\sigma}^{\SUBR}(\omega,T) -
\im\tilde{\Sigma}_{\sigma}^{\SUBI}(\omega,T)$); satisfying
\begin{equation}
\tilde{\Sigma}_{\sigma}(\omega,T) =
-[\tilde{\Sigma}_{-\sigma}(-\omega,T)]^{*}
\end{equation}
for the p-h symmetric case considered. It is precisely this
two-self-energy framework that underlies the LMA
\cite{lg00,gl03,gl032,let,dl01,gl02,ld02,ld01a,ldb01} (with the single 
self-energy
obtained if desired as a byproduct, following simply from direct
comparison of equations (2.10, 
11) with equation (2.2)); requisite
details will be given in section 4. The LM phase is then characterised
by \emph{non-vanishing}, spin-dependent `renormalized levels'
$\tilde{\Sigma}^{\SUBR}_{\sigma}(0,0) =
-\tilde{\Sigma}^{\SUBR}_{-\sigma}(0,0)$ (so called because they
correspond to $\tilde{\epsilon}_{\ii\sigma} = \epsilon_{\ii} +
\tilde{\Sigma}^{\SUBR}_{\sigma}(0,0)$ if the Hartree contribution
is explicitly retained in the
$\tilde{\Sigma}^{\SUBR}_{\sigma}(\omega,0)$). From this, using
equations (2.10,11), the low-$\omega$ behaviour embodied in
equation (2.8) (and hence equation (2.9) for the single
self-energy) \emph{follows directly}, namely
\begin{equation}
\pi D(\omega,0) \stackrel{|\omega| \rightarrow 0}{\propto}
|\omega|^{r}/ [\tilde{\Sigma}^{\SUBR}_{\sigma}(0,0)]^{2}
\end{equation}
(assuming merely that $\tilde{\Sigma}^{\SUBI}_{\sigma}(\omega,0)$
vanishes no more rapidly than the hybridization $\propto
|\omega|^{r}$). As $U \rightarrow U_{\cc}+$ and the
LM$\rightarrow$GFL transition is approached, the renormalized
level $\tilde{\Sigma}^{\SUBR}_{\sigma}(0,0) \rightarrow 0$ 
\cite{gl03,gl032}; and
$\tilde{\Sigma}^{\SUBR}_{\sigma}(0,0) =0$ remains throughout the
(`symmetry unbroken') GFL phase $U < U_{\cc}$  \cite{lg00,gl03,gl032},
such that the characteristic low-$\omega$ spectral behaviour equation
(2.5) is likewise correctly recovered. 

  The underlying two-self-energy description is thus
capable of handling both phases simultaneously, and hence the
transition between them. Further, since the renormalized level
$|\tilde{\Sigma}^{\SUBR}_{\sigma}(0,0)|$ vanishes as the
LM$\rightarrow$GFL transition is approached, it naturally generates a
low-energy scale $\wl$ characteristic of the LM phase (defined precisely 
in section 5.2); in terms of
which LM phase dynamics in the vicinity of the transition have
recently been shown \cite{gl03,gl032} to exhibit universal scaling.
This behaviour is of course the counterpart of the Kondo scale
$\wk$ and associated spectral scaling in the GFL phase, but now for
the LM phase on the other side of the underlying quantum
phase transition.

\seceq
\section{Scaling: general considerations}

  Close to and on either side of the quantum phase transition, the problem
is thus characterised by a single low-energy scale, denoted
generically by $\omega_{*}$ (the Kondo scale for the GFL phase;
$\wl$ for the LM phase). Given this, and without recourse to any
particular theory, we now show that general scaling arguments may
be used to obtain a number of exact results for the finite-$T$ scaling
spectra, in both the GFL and LM phases \it and at the QCP itself
\rm (where $\omega_{*} =0$ identically). These are important in
themselves, and in addition provide stringent dictates that should
be satisfied by any approximate theoretical approach to the
problem.

  We consider first the case of $T=0$, for which general
scaling arguments made by us in \cite{gl032} serve as a starting
point. In the following  we denote the spectrum
by $D(U;\omega)$, with the $U$-dependence temporarily
explicit. As the
transition is approached,  $u =|1-U/U_{\cc}(r)| \rightarrow 0$,
the
low-energy scale $\omega_{*}$ vanishes, as
\begin{equation}
\omega_{*} = u^a
\end{equation}
with exponent $a$. $D(U;\omega)$ can now be expressed generally in
the scaling form $\pi D(U;\omega) =
u^{-ab}\Psi_{\alpha}(\omega/u^a)$ in terms of two exponents $a$
and $b$ (and with $\alpha =$ GFL or LM denoting the appropriate
phase), i.e.\ as
\begin{equation}
\pi\omega_{*}^{b}D(U;\omega) = \Psi_{\alpha}(\omega/\omega_{*})
\end{equation}
with the exponent $a$ eliminated and the $\omega$-dependence encoded solely
in $\omega/\omega_{*} = \tilde{\omega}$. Equation (3.2) simply embodies the
universal scaling behaviour
of the $T=0$ single-particle spectrum close to the QPT. Using it, three results
follow: 

(i) That the exponent $b=r$ for the GFL phase (and for all $r$
where the GFL phase arises). This follows directly from equation
(3.2) using the fact that the leading $\omega \rightarrow 0$
behaviour of $D(U;\omega)$ throughout the GFL phase is given
exactly by equation (2.5), together with the fact that
$\Psi_{\alpha}(x)$ is universal; i.e.\
\begin{equation}
\pi\omega_{*}^{r}D(U;\omega) = \Psi_{\alpha}(\omega/\omega_{*}).
\end{equation}

(ii) Now consider the approach to the QCP, $\omega_{*} \rightarrow 0$ and hence
$\omega/\omega_{*} \rightarrow \infty$. Since the QCP itself must be `scale free'
(independent of $\omega_{*}$), equation (3.3) implies
\begin{equation}
\Psi_{\alpha}(\omega/\omega_{*}) \stackrel{|\omega|/\omega_{*}\rightarrow \infty}
{\sim} C(r)(|\omega|/\omega_{*})^{-r}
\end{equation}
with $C(r)$ a constant (naturally $r$-dependent). From equations (3.3,4) it follows
directly that precisely at the QCP ($\omega_{*} =0$),
\begin{equation}
\pi D(U_{\cc};\omega) = C(r)|\omega|^{-r}
\end{equation}
which gives explicitly the $\omega$-dependence of the ($T=0$) QCP spectrum.

(iii) Assuming naturally that the QCP behaviour is independent of the phase
from which it is approached, equation (3.4) applies also to the LM phase. From
equations (3.2) and (3.4) it follows that the exponent $b=r$ for the LM phase
as well (as one might expect physically).

  The above behaviour is indeed as found in practice from the
LMA at $T=0$ \cite{lg00,gl03,gl032}, with the scaling for
equation (3.3) also confirmed by NRG calculations \cite{bglp}. But
the important point here is that these results are general,
independent of approximations (be they analytical or numerical),
and as such holding across the entire $r$-range for which the symmetric
QCP exists, i.e.\ $0<r< \frac{1}{2}$ \cite{gbi98}.
We add morover that while the argument above for $b=r$ leading to
equation (3.3) is particular to the symmetric PAIM considered
explicitly here,
recent LMA results for the asymmetric PAIM for $0<r<1$ \cite{gl032}, and
NRG results \cite{vb02} for the asymmetric pseudogap Kondo
model for $\frac{1}{2} < r < 1$, are also consistent with the
exponent $b=r$.  The QCP behaviour equation (3.5) then
follows directly, and the LMA/NRG results of Ref.\ \onlinecite{vb02,gl032} are
likewise consistent with it.

  The arguments above \cite{gl032} can now be extended to finite
temperature. The scaling
spectra may again be cast generally in the form $\pi
D(U;\omega,T)
= u^{-ar}\Psi_{\alpha}(\omega/u^a, T/u^a)$, i.e.\ as
\begin{equation}
\pi \omega_{*}^{r}D(U;\omega,T) =
\Psi_{\alpha}(\omega/\omega_{*},T/\omega_{*})
\end{equation}
expressing the fact that $\omega_{*}^{r}D(U;\omega,T)$ is
a universal function of $\tilde{\omega} =\omega/\omega_{*}$ \it
and \rm $\tilde{T} = T/\omega_{*}$.  From this follow three results,
new to our knowledge:-

(i) Consider first the approach to the QCP at finite-$T$, i.e.\
$\tilde{\omega} \rightarrow \infty$ and $\tilde{T} \rightarrow
\infty$ such that $\tilde{\omega}/\tilde{T} = \omega/T$ is fixed.
Again, since the QCP must be scale free,
$\Psi_{\alpha}(\tilde{\omega} \rightarrow \infty, \tilde{T}
\rightarrow \infty)$ must be of form $\tilde{T}^{p}S(\omega/T)$
with the exponent $p=-r$ from equation (3.6), i.e.\
\begin{equation}
\Psi_{\alpha}(\tilde{\omega},\tilde{T}) \sim \tilde{T}^{-r}S(\omega/T).
\end{equation}

(ii) From equations (3.6,7) it follows directly that precisely at the
QCP ($\omega_{*} =0$):
\begin{equation}
\pi T^{r} D(U_{\cc};\omega,T) = S(\omega/T)
\end{equation}
i.e.\ $T^{r}D(U_{\cc};\omega,T)$ exhibits \it pure
$\omega/T$-scaling\rm.

(iii) The large-$x$ behaviour of $S(x)$ --- and hence the `tail'
behaviour of the QCP scaling spectrum $S(\omega/T)$ --- may be
deduced on the natural assumption that the limits  $\omega_{*}
\rightarrow 0$ and $T \rightarrow 0$ commute; i.e.\ that the $T=0$
QCP spectrum may either be obtained at $T=0$ from the limit
$\omega_{*} \rightarrow 0$ (as in equation (3.5)), or from the $T
\rightarrow 0$ limit of the finite-$T$ QCP scaling spectrum
equation (3.8) (in which $\omega_{*} \rightarrow 0$ has been taken
first). With this, comparison of equations (3.5,8) yields the
desired result
\begin{equation}
S(\omega/T) \stackrel{|\omega|/T \rightarrow \infty}{\sim}
C(r)(|\omega|/T)^{-r}.
\end{equation}

  The behaviour embodied in equations (3.7--9) follows
on general grounds. In particular, the $\omega/T$-scaling of the
QCP scaling spectrum is a key conclusion. Neither is such scaling
specific to single-particle dynamics. For the dynamical local
magnetic susceptibility, recent NRG calculations \cite{ingsi02} are
consistent with $\omega/T$-scaling at the QCP (for $0<r<1$), as
supported further by approximate analytic results for small $r$
based on a procedure analogous to the standard
$\epsilon$-expansion. The preceding arguments do \emph{not} of course
determine the functional form of the scaling spectra embodied in
the $\Psi_{\alpha}(\tilde{\omega},\tilde{T})$ and $S(\omega/T)$,
save for their asymptotic form equations (3.4,9): for that a
`real' (inevitably approximate) theory is required, as considered
in the following sections; and from which the general results
above should of course be recovered.
  We note in passing that the  
arguments above also encompass the role of a local
magnetic field, $h = \frac{1}{2}g\mu_{\mbox{\ssz{B}}}H_{\mbox{\ssz{loc}}}$ (with $\tilde{h}
= h/\omega_{*}$ in the following): replacing $T$ by $h$ in
equations (3.6-9) gives the appropriate scaling behaviour in terms
of $\tilde{\omega}$ and $\tilde{h}$ (for $T=0$). At the QCP in
particular $h^{r}D(U_{\cc};\omega,h)$ exhibits \it pure
$\omega/h$-scaling\rm,
\begin{equation}
\pi h^{r}D(U_{\cc};\omega,h) = S^{\prime}(\omega/h)
\end{equation}
with the large-$|\omega|/h$ `tail' behaviour  $S^{\prime}(\omega/h)
\sim C(r)(|\omega|/h)^{-r}$.

\subsection{The low-energy scale}

  Before proceeding we return briefly to an obvious issue: the physical
nature of the low-energy scale $\omega_{*}$.  In the GFL phase for
example, how is it related to the quasiparticle weight $Z
=[1-(\partial\Sigma^{\SUBR}(\omega;0)/
\partial\omega)_{\omega=0}]^{-1}$, which we likewise expect to vanish as $U \rightarrow
U_{\cc}(r)-$,
\begin{equation}
Z \sim u^{a_{1}}
\end{equation}
with exponent $a_{1}$? That question may be answered most simply
(albeit partially here) by considering the limiting   low-frequency
`quasiparticle form' for the impurity Green function, whose
behaviour reflects the adiabatic continuity to the non-interacting
limit that is intrinsic to the GFL phase. This follows from the
low-$\omega$ expansion of the single self-energy,
$\Sigma(\omega;0) \sim -(\frac{1}{Z}-1)\omega$ (with the imaginary
part neglected as usual \cite{hew}); using which with equations
(2.2--4) yields the quasiparticle behaviour
\begin{equation}
\pi D(U;\omega) \sim \frac{|\omega|^{r}} {(\frac{\omega}{Z} +
\mbox{sgn}(\omega)\beta(r)|\omega|^{r})^{2}+ |\omega|^{2r}}
\end{equation}
which itself must be expressible in the scaling form equation (3.3). This in
turn implies that $\omega_{*}$ and $Z$ are related by
\begin{equation}
\omega_{*} = [k_{1}Z]^{\frac{1}{1-r}}
\end{equation}
(with $k_{1}$ an arbitrary constant up to which low-energy scales are defined),
such that equation (3.12) reduces to the required form
\begin{equation}
\pi \omega_{*}^{r}D(U;\omega) \sim \frac{|\tilde{\omega}|^{r}}
{(k_{1}|\tilde{\omega}| +
\mbox{sgn}(\omega)\beta(r)|\tilde{\omega}|^{r})^{2} +
|\tilde{\omega}|^{2r}}
\end{equation}
with $\tilde{\omega} = \omega/\omega_{*}$. For the GFL phase the
$\omega_{*}$ ($\equiv \omega_{K}$) scale and the quasiparticle
weight $Z$ are thus related by equation (3.13) (and hence the
exponents $a$ (equation (3.1)) and $a_{1}$ (equation (3.11)) by $
a_{1} = a(1-r))$. That behaviour is correctly recovered by the
LMA, which in addition shows $\omega_{*}$ to be equivalently the
characteristic Kondo spin-flip scale associated with transverse
spin excitations; see section 5.1 below.

  In contrast to the GFL phase where the leading low-$\omega$ spectral behaviour
is $D(U;\omega) \propto |\omega|^{-r}$ as above, the corresponding
behaviour for the LM phase is given by $\pi D(U;\omega) \propto
|\omega|^{r}/ [\tilde{\Sigma}^ 
{\SUBR}_{\sigma}(0;0)]^{2}$
(equation (2.13)) in terms of the `renormalized level'
$\tilde{\Sigma}^{\SUBR}_{\sigma}(0;0)$, which is non-zero in the
LM phase and vanishes as $U \rightarrow U_{\cc}(r)+$ \cite{gl03,gl032};
i.e.\
\begin{equation}
|\tilde{\Sigma}^{\SUBR}_{\sigma}(0;0)| \sim u^{a_{2}}
\end{equation}
with exponent $a_{2}$. Since equation (2.13) must be expressible in the scaling
form equation (3.3) it follows that $\omega_{*} (\equiv \omega_{L})$ in the LM
phase is related simply to the renormalized level by
\begin{equation}
\omega_{*} =
[k_{2}|\tilde{\Sigma}^{\SUBR}_{\sigma}(0;0)|]^{\frac{1}{r}}
\end{equation}
(and hence the exponents $a$ and $a_{2}$ by $a_{2}=ar$). We emphasize again
that the exponents obtained above are exact, valid for all $0<r<\frac{1}{2}$.

 After the above general considerations we now consider the
finite-temperature scaling properties of the PAIM within the local moment
approach.

\seceq
\section{Local moment approach: finite $T$}
Regardless of the phase considered the LMA uses the two-self energy
description embodied in equations (2.10,11), with the self-energies
separated for convenience as \be \sts(\w^+,
T)=-\frac{\sigma}{2}U|\mu|+\Sigma_{\sigma}(\w^+, T) \ee into a
purely static Fock piece with local moment $|\mu|$ (that is
retained alone at mean-field level); and an all important dynamical contribution
$\Sigma_{\sigma}(\w^+,T)$ containing the spin-flip physics that
dominates at low-energies.  The $G_{\sigma}(\w,T)$ (equation (2.11)) may be recast equivalently as \be
G_{\sigma}(\w,T)=\left[\scrG_{\sigma}(\w)^{-1}-\Sigma_{\sigma}(\w^+,T)\right]^{-1}
\ee in terms of the mean-field (MF) propagators \be
\scrG_{\sigma}(\w)=[\w^++\frac{\sigma}{2}U|\mu|-\Delta(\w)]^{-1}
\ee and with $\Sigma_{\sigma}\equiv \Sigma_{\sigma}[\{\scrG_{\sigma}\}]$ a
functional of the $\{\scrG_{\sigma}(\w)\}$.

\begin{figure}
\resizebox{0.3\columnwidth}{!}{%
  \includegraphics{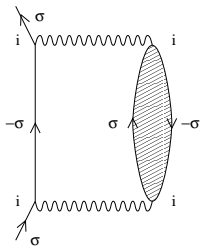}}
\caption{Standard LMA diagram for the dynamical contribution  
 $\Sigma_{\sigma}$ to the self-energies.}
\label{fig:1}       
\end{figure}

The standard LMA diagrammatic approximation for the $t$-ordered
two-self-energies is depicted in figure 1.  For full details, including
their physical interpretation, the reader is referred to
\cite{lg00,let,gl02}. In retarded form at finite temperature, $\Sigma_{\up}(\w)$
is given by \cite{ld02} \bea
\Sigma_{\up}(\w^+,T)&=&U^2\int_{-\infty}^{\infty}\frac{\rmd
\w_1}{\pi}\ \int_{-\infty}^{\infty}\rmd \w_2\
\chi^{+-}(\w_1,T) \nonumber \\
&\times& \frac{D^0_{\down}(\w_2)}{\w+\w_1-\w_2+\im\eta}\ g(\w_1;
\w_2) \eea where
$D^0_{\sigma}(\w)=-\frac{1}{\pi}\mbox{Im}\scrG_{\sigma}(\w)$ is the MF
spectral density; and $g(\w_1;\w_2)=\theta(-\w_1)$  $[1-f(\w_2;
T)]+\theta(\w_1)f(\w_2; T)$ with $f(\w; T)
=[1+\mbox{exp}(\w/T)]^{-1}$ the Fermi function.  
$\chi^{+-}(\w,T)\equiv \mbox{Im}\Pi^{+-}(\w,T)$
 is the spectrum of transverse spin excitations, with $\Pi^{+-}(\w,T)$ 
 the finite-$T$ polarization propagator, calculated via  an RPA-like p-h 
 ladder sum \cite{ld02}.  $\Sigma_{\down}(\w)=-[\Sigma_{\up}(-\w)]^{*}$ follows
trivially by
particle-hole symmetry;  we thus choose to work with
$\Sigma_{\up}(\w)$.

The $T$-dependence of the $\chi^{+-}(\w,T)$, and also of 
the local moment $|\mu|=|\mu(T)|$ entering the MF propagators
(equation(4.3)) and self-energies (equation (4.1)), 
 is found to be insignificantly
small, provided we remain uninterested in temperatures on the order
of the non-universal energy scales in the problem.
This is precisely as for the $r=0$ case \cite{ld02} and we
thus work with $\chi^{+-}(\w,T)\simeq\chi^{+-}(\w,T=0)
\equiv\mbox{Im}\Pi^{+-}(\w)$ and local moment $|\mu(0)|\equiv |\mu|$.  The
remaining $T$-dependence, describing the universal scaling regime, is controlled
solely by $g(\w_1; \w_2)$ entering equation (4.4).

In describing the GFL phase within the LMA, the key concept is that
of symmetry restoration (SR) \cite{lg00,gl032,let,gl02}: self-consistent
restoration of the symmetry broken at pure MF level. In mathematical terms
this is encoded simply in  $\st_{\up}(0^+,T=0)
=\st_{\down}(0^+,T=0)$ and hence, via particle-hole symmetry,
\be \left[\st_{\up}(0^+; T=0 )\equiv\right]\
\Sigma_{\up}(0)-\frac{1}{2}U|\mu|=0.\ee  Equation (4.5) guarantees in particular 
the correct Fermi
liquid form ${\cal F}(0,0)=1$ (see equation (2.7)). It is
satisfied in practice, for any given $U<U_{c}$, by varying the local moment
$|\mu|$ entering the MF propagators from the pure MF value $|\mu_0|<|\mu|$
until equation (4.5) --- a single condition at the Fermi level $\omega =0$
--- is satisfied.
 In so doing a low-energy
scale is introduced into the problem through a strong resonance in
$\mbox{Im}\Pi^{+-}(\w)$ centred on $\w =
\wm \equiv \wm(r)$.  This is the Kondo or spin-flip scale $\wm
\propto \wk$ and sets the timescale for symmetry restoration,
$\tau \sim h/\wm$ \cite{lg00,gl032,let,gl02}.

In the LM phase, by contrast, it is not possible to satisfy the
symmetry restoration condition equation (4.5) and $|\mu|=|\mu_0|$
\cite{lg00,gl032}:
$\mbox{Im}\Pi^{+-}(\w)$ then correctly contains a delta-function contribution at
$\w = 0$, reflecting physically the zero-energy cost for spin-flips in the
doubly-degenerate LM phase.  The `renormalized levels' (see section 2.2)
$\{\sts(0,0)\}$ are here found to be non-zero and sign-definite.

The GFL/LM phase boundary, i.e. $U_{\cc} \equiv {U}_{\cc}(r)$, may then be accessed
either as the limit of solutions to equation (4.5), where $\wm\ra
0$; or, coming from the LM side, as the limiting ${U}$ for which the
$\{\sts(0,0)\}$ vanish.  In either case the \emph{same} ${U}_{\cc}(r)$
obtains; and both GFL and LM phases are correctly found to arise for all
$0<r<\frac{1}{2}$, while solely LM states occur for all $r>\frac{1}{2}$
and ${U}>0$ \cite{lg00}. For $r \rightarrow 0$ the LMA recovers the
exact result that $U_{\cc}=8/\pi r$ which, together with the low-$r$ result
$a=1/r$ (see equation (3.1)) \cite{lg00} recovers the exact exponential
dependence of the regular $r=0$ metallic AIM \cite{hew}. 

The practical strategy for solving the problem at finite-$T$ is likewise
straightforward.  Once the local moment $|\mu|$ is known from symmetry
restoration
(for any given $r$ and $U$), the dynamical self-energy
$\Sigma_{\up}(\w^{+}, T)$ follows from equation (4.4).  The full
self-energy then follows immediately from equation (4.1) and $G(\w,T)$ in turn via
equations (4.2) and (2.10).
We now give numerical
results obtained in this way for the GFL phase at finite-$T$.

\subsection{Numerical results: GFL phase scaling}
Here we show that the LMA correctly captures scaling of
the single-particle spectrum $D(\w,T)$ as the transition is
approached, $U\ra U_{\cc}(r)-$.  That is, $\wm^rD(\w,T)$
scales in terms of both $\tilde{\w}=\w/\wm$ and $\tilde{T}=T/\wm$,
as expressed in equation (3.6);  or equivalently that
${\cal F}(\w, T)\equiv F(\tilde{\w},\tilde{T})$.

\begin{figure}
\begin{center}
\psfrag{xaxis1}[bc][bc]{$\wt$}
\psfrag{yaxis1}[bc][bc]{${\cal F}(\w, T)$}
\psfrag{xaxis2}[bc][bc]{$\w$}
\epsfig{file=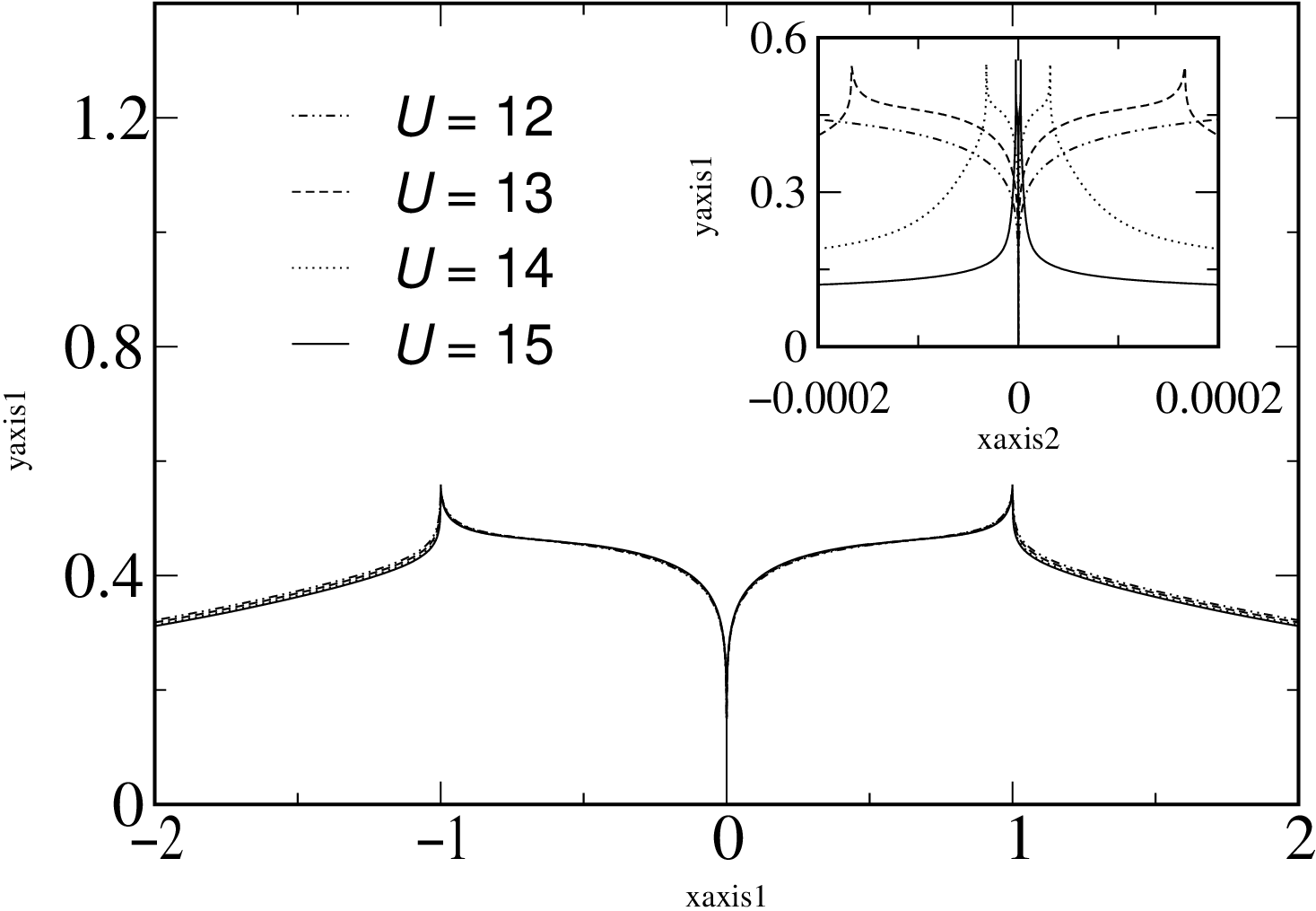, width=5cm, angle=270} 
\caption{GFL phase scaling spectrum ${\cal
F}(\w,T)=\pi\mbox{sec}^2(\frac{\pi}{2}r)|\w|^rD(\w,T)$ versus
$\tilde{\w}=\w/\wm$, for fixed $\tilde{T}=T/\wm=1$ and four values
of the interaction strength $U$ approaching $U_{\cc}\simeq 16.4$.
The inset shows ${\cal F}(\w,T)$ on an absolute scale.}
\end{center}
\end{figure}

Figure 2 shows the numerically determined spectral function ${\cal F}(\w,T)$
versus $\tilde{\w}=\w/\wm$ for $r=0.2$, a fixed temperature
$\tilde{T}=T/\wm=1$ and four values of ${U}<{U}_{\cc}(r)$ as the
transition is approached (${U_{\cc}}\simeq 16.4$).  The spectra
--- very different on an absolute scale as shown in the inset to
the figure --- indeed clearly collapse to a common scaling form as ${U}\ra
{U}_{\cc}(r)-$ and $\wm$ becomes arbitrarily small.  

This
behaviour is not of course
particular to the choice $\tilde{T}=1$, but rather is found to
arise for all (finite) $\tilde{T}$.
 Figure 3 shows the resultant scaling spectra for a representative range of 
 $\tilde{T}$, labelled in the figure.
   For $T=0$ the GFL phase
 Kondo resonance is fully present and ${\cal F}(\w=0,T=0)=1$, as
 discussed in section 3.  As temperature is raised  through a range
 on the order of the Kondo scale itself ($\tilde{T}\sim
 \Or(1)$) the Kondo resonance is thermally destroyed.  First it is
 split --- reflecting the fact that the Fermi level $D(\w =0,T)$ is finite
 for $T>0$ (see inset) --- and then progressively eroded as $\tilde{T}$ 
 increases.  For any given temperature $\tilde{T}$ the principal effect
 of temperature
 arises on frequency scales $|\tilde{\w}|\lesssim \tilde{T}$, such
 that for $\tilde{\w}\gg \tilde{T}$ the spectrum is essentially
 coincident with the $\tilde{T}=0$ limit.  This behaviour is
 observed clearly in the inset to figure 3 wherein e.g.\ the $\tilde{T}=1$
 scaling spectrum is coincident with that for $\tilde{T}=0$ for
 $|\tilde{\w}|\gtrsim 10$; and that for $\tilde{T}=10$ coincides
 with $\tilde{T}=0$ for $|\tilde{\w}|\gtrsim 50$. We also add that the small
spectral feature at $\wt \simeq 1$, seen e.g.\ in the inset to figure 3,
is entirely an artefact of the specific RPA form for Im$\Pi^{+-}(\w)$ that,
as discussed in detail in previous work \cite{dl01}, can be removed entirely
with both little effect on the appearance of the spectrum and no effect on any
asymptotic results.

\begin{figure}
\begin{center}
\psfrag{xaxis}[bc][bc]{$\wt$}
\psfrag{yaxis}[bc][bc]{${\cal F}(\w, T)$}
\psfrag{xlabel}[bc][bc]{$\wt$}
\psfrag{ylabel}[bc][bc]{\hspace{0.5cm}${\wm}^rD(\w, T)$}
\psfrag{t0}[bl][bl]{$\tilde{T} = 0$}
\psfrag{t0.1}[bl][bl]{$\tilde{T}=0.1$}
\psfrag{t1}[bl][bl]{$\tilde{T}=1$}
\psfrag{t10}[bl][bl]{$\tilde{T}=10$}
\psfrag{xxxt100}[bl][bl]{$\tilde{T}=100$}
\epsfig{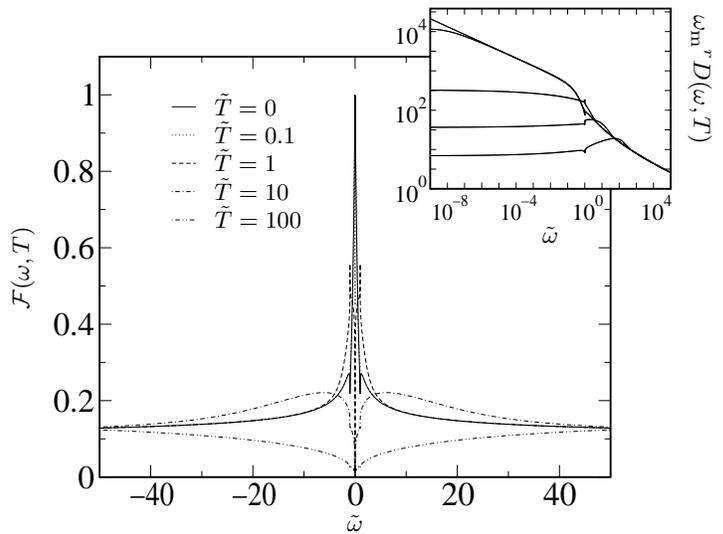} \caption{Numerical results: GFL phase scaling
spectrum ${\cal F}(\w,T)$ versus $\wt=\w/\wm$ for a range of
$\tilde{T}=T/\wm$.  The generalized Kondo resonance is thermally
destroyed, with the principal effects of temperature arising for
$|\wt| \lesssim \tilde{T}$.  The inset shows $\wm^rD(\w,T)$ versus
$\wt$ on a logarithmic scale for the same $\tilde{T}$, increasing top 
to bottom; see text for comments. }
\end{center}
\end{figure}

\seceq
\section{Results}
We now turn to an analytical description of the finite-$T$ scaling
spectrum in the strong coupling ($=$ strongly correlated) limit of ${U}\gg 1$.  As in
Ref.\ \onlinecite{gl03}, and in contrast to the results of section 3,  this formally restricts discussion to $r\ll 1$, although as demonstrated below such an analysis does rather well in accounting for the numerical results given in section 4.1 for
$r=0.2$.  For this reason we present only analytical results in
the remainder of the paper, where we focus in turn on the GFL phase, LM phase 
and the QCP itself.

\subsection{GFL phase} To obtain the scaling behaviour of $D(\w)$
we consider finite $\tilde{\w} =\w/\w_{m}$ in the formal limit $\wm\ra 0$,  
thereby projecting out irrelevant non-universal spectral features (e.g.\ the
Hubbard bands at $\omega \sim \pm U/2$). 
 Using equations (2.10,11) the scaling
spectrum $\wm^rD(\w,T)$ is thus given
by \be
 \wm^rD(\w,T)=-\frac{1}{2\pi}\mbox{Im}\sum_{\sigma}\left[( \wm^r)^{-1}
(\Delta(\w)-\sts(\w,T)\right]^{-1} \ee 
where (from equations (2.3,4)) the hybridization
 $\Delta(\w)=\Delta(\tilde{\w}\wm)$ reduces simply to
 $( \wm^r)^{-1}\Delta(\w)=-\mbox{sgn}(\w)[\beta(r)+\im]|\wt|^r$
(and, trivially, the bare $\w=\wm\wt$ of equation (2.11) may be neglected). 
We consider explicitly $\w \geq 0$ in the following, since
$D(-\w,T)=D(\w,T)$ from particle-hole symmetry.

Our task now is to demonstrate that $( \wm^r)^{-1}\st_{\up}(\w,T)$ is a
function solely of $\wt$ and $\Tt$ in the GFL phase.  
It follows from equation (4.4) that 
\bea
\Sigma_{\up}^{\subi}(\w,T)&=& U^2\int^{\infty}_{-\infty}\rmd \w_1\
\chi^{+-}(\w_1)D^0_{\down}(\w_1+\w) \\
&\times&\left[\theta(\w_1)f(\w_1+\w)+\theta(-\w_1)(1-f(\w_1+\w))\right]
\nonumber
\eea
(with the temperature dependence of $f(\w,T)$ temporarily
omitted).
In the requisite strong coupling regime $U\gg 1$,
$\chi^{+-}(\w,T)\simeq\mbox{Im}\Pi^{+-}(\w)=\pi\delta(\w-\wm)$ is readily shown
to reduce asymptotically to a $\delta$-function centred
on $\w =\w_{m}$; and
thus $\Sigma_{\up}^{\subi}(\w,T)=\pi U^2
D_{\down}^0(\wm[1+\wt])f(1+\wt,\Tt)$ (noting
that $f(\w,T)$ depends solely on the ratio $\w/T$).  
Using equation (4.3), $D_{\downarrow}^{0}(\omega) \sim 4|\w|^{r}/\pi U^{2}$
in strong coupling;
whence \be (
\wm^r)^{-1}\Sigma_{\up}^{\subi}(\w,T)=4|1+\wt|^rf(1+\wt,\Tt)\ee
which is indeed universally dependent solely on $\wt$ and $\Tt$.
We now consider the real part $\st^{\subr}_{\up}(\w,T)$.  For $T=0$ it is 
known to be given by \cite{gl03} \be (
\wm^r)^{-1}\st_{\up}^{\subr}(\w,0)=-\gamma(r)[|1+\wt|^r-1]\ee
where $\gamma(r)=4/\mbox{sin}(\pi r)\sim 4/\pi r$, scaling
solely in terms of $\wt$ and correctly satisfying symmetry restoration
($\st_{\up}^{\subr}(0,0)=0$) for the GFL phase.  The finite-$T$
difference
$\delta\st^{\subr}_{\up}(\w,T)=\Sigma^{\subr}_{\up}(\w,T)-\Sigma^{\subr}_{\up}(\w,0)$
is readily calculated via Hilbert transforms using
equation (5.2).  After some manipulation it is found to be given by \be (
\wm^r)^{-1}\delta\st^{\subr}_{\up}(\w,T)=-\frac{4}{\pi}\Tt^r
H\left(\frac{|1+\wt|}{\Tt}\right).\ee $H(y)$ is 
defined as \be H(y)=\int^{\infty}_0\rmd\w\
\frac{\w^r}{\mbox{exp}(\w)+1}\ {\cal P}\left(
\frac{1}{\w+y}+\frac{1}{\w-y}\right)\ee  
and is trivial to evaluate numerically; its $y>>1$ and $y<<1$ asymptotic
behaviour is also obtainable in closed form, as used below. 
From equations (5.4,5),
we thus obtain \bea
( \wm^r)^{-1}\st^{\subr}_{\up}(\w,T)=&-&\gamma(r)[|1+\wt|^r-1]\nonumber\\
&-&\frac{4}{\pi}\Tt^r H\left(\frac{|1+\wt|}{\Tt}\right).\eea

The complete GFL scaling spectrum now follows from equations (5.1)
and (5.3,7); and, as required, $ \wm^rD(\w,T)$ is seen to be a universal
function of $\wt =\w/\wm$ and $\Tt =T/\wm$. 

Before proceeding to a comparison with
figure 3 however,  we add that $ \wm^rD(\w,T)$ is
equivalently a universal function of $\w/T\equiv
\w'$ and $\Tt^{-1}$. To see this note that
equations (5.3,7) may be rewritten as \be (
T^r)^{-1}\Sigma_{\up}^{\subi}(\w,T)=4\left|\w'+\frac{1}{\Tt}\right|^rf\left(
\frac{1}{\Tt}+\w'\right)\ee and \bea (
T^r)^{-1}\st^{\subr}_{\up}(\w,T)=&-&\gamma(r)\left[\left|\w'+\frac{1}{\wt}\right|^r-
\Tt^{-r}\right]\nonumber\\
&-&\frac{4}{\pi}H\left(\left|\w'+\frac{1}{\Tt}\right|\right)\eea
(with notation $f(z)\equiv f(z,1)=[\mbox{exp}(z)+1]^{-1}$ for the Fermi
functions), which depend solely on $\w'$ and $\Tt^{-1}$.
From equation (5.1) it follows directly that
\be \pi T^rD(\w,T)=-\frac{1}{2}\mbox{Im}\sum_{\sigma}\left\{( T^r)^{-1}
\left[\Delta(\w)-\sts(\w,T)\right]\right\}^{-1} \ee 
which is thus of form 
\be
 \pi T^rD(\w,T)\equiv
S_{\mbox{\ssz{GFL}}}\left(\frac{\w}{T},\frac{1}{\Tt}\right). \ee
The importance of this result is that from it the finite-$T$ spectrum
\emph{at the QCP itself} (\emph{cf} equation (3.8)) may be
obtained explicitly, since the QCP corresponds 
to $\wm = 0$ i.e.\ $\Tt^{-1}=0$. We consider this issue
explicitly in section 5.3.

\begin{figure}
\begin{center}
\psfrag{xaxis}[bc][bc]{$\wt$}
\psfrag{yaxis}[bc][bc]{${\cal F}(\w, T)$}
\psfrag{t1}[bl][bl]{$\tilde{T}=0.1$}
\psfrag{t2}[bl][bl]{$\tilde{T}=1$}
\psfrag{t3}[bl][bl]{$\tilde{T}=10$}
\psfrag{t4}[bl][bl]{$\tilde{T}=100$}
\psfrag{t0}[bl][bl]{$\tilde{T}=0$}

\epsfig{file=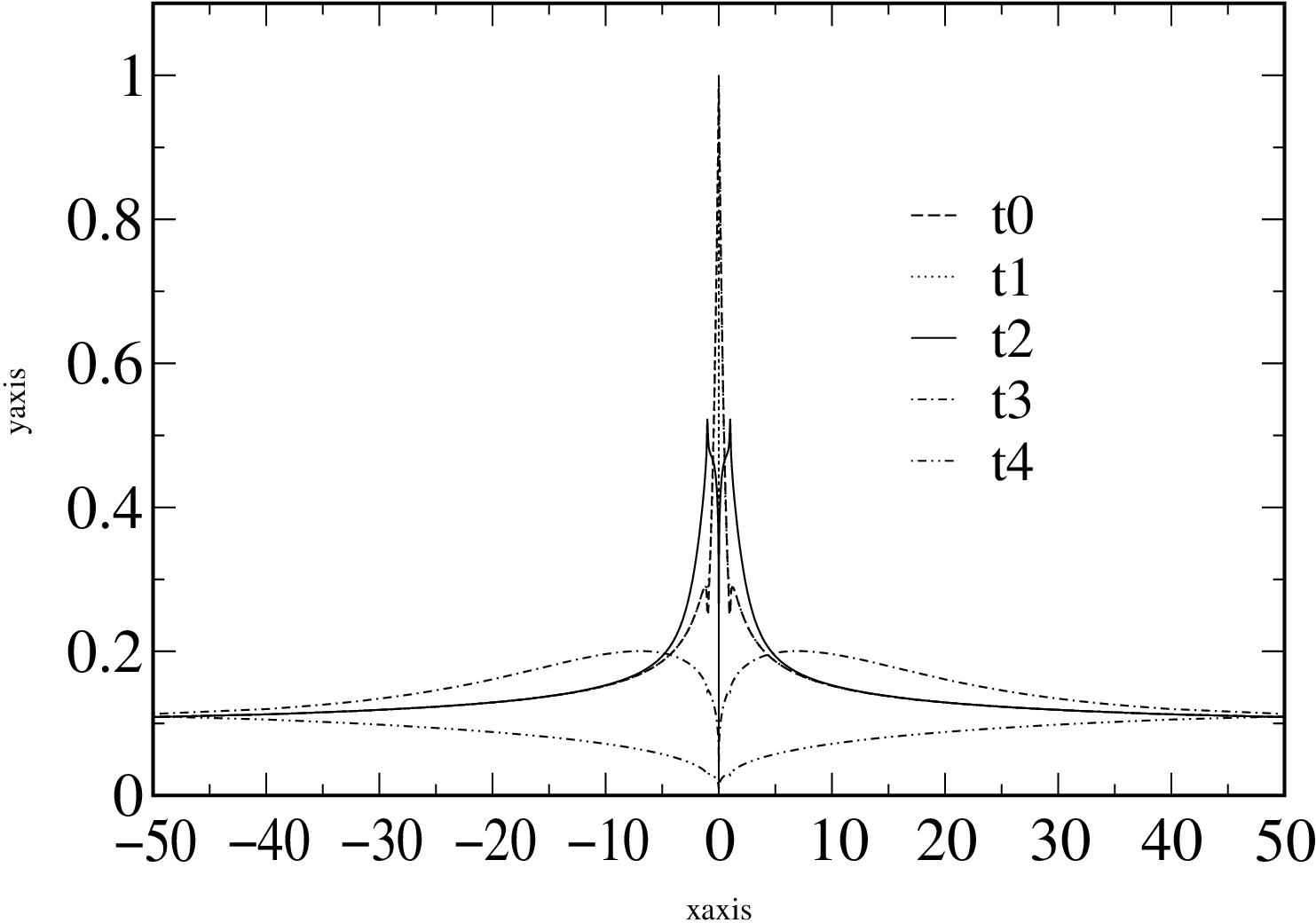,width=6cm, angle=270} \caption{Analytical results: GFL
phase scaling spectrum ${\cal F}(\w,T)$ versus $\wt=\w/\wm$ for
$r=0.2$ and the sequence of temperatures shown.}
\end{center}
\end{figure}

First we compare directly the analytic  results above for
the GFL phase, with the numerical results given in figure 3. Figure
4 shows the modified spectral function ${\cal F}(\w)$ versus
$\wt=\w/\wm$ for $r=0.2$ arising from equation (5.1,3,7), for 
the same range of `scaled' temperature $\Tt=T/\wm$.
The agreement is seen to be excellent: all features
of the numerical spectra are captured by the analytical form; quantitative
differences are small, and arise 
because the equations hold asymptotically as $r\ra 0$.

Two specific points concerning the large-$\wt$ behaviour of the
GFL phase scaling spectrum may now be made.  First,
the observation (section 4.1) that for frequencies
$|\wt|\gg\Tt$ the scaling spectra are effectively independent of
$\Tt$.  For $|\wt|\gg\mbox{max}(1,\Tt)$ the argument of
$H(|1+\wt|/\Tt)$ is large and $H(|1+\wt|/\Tt)\simeq 0$ may be
neglected in equation (5.7) ($H(y)\sim -\pi^2/6y^2$ for $y\gg
1$).  Hence $( \wm^r)^{-1}\st_{\up}^{\subr}(\w,T)\sim
-\gamma(r)[|\wt|^r-1]$ and $(
\wm^r)^{-1}\st_{\up}^{\subi}(\w,T)\sim 4|\wt|^r\theta(-[1+\wt])$
from equations (5.7,3), 
and from equation (5.1) it follows that 
\bea \pi
\wm^rD(\w,T)&\sim&\frac{1}{2|\wt|^r}\left\{\frac{1}{[\beta(r)+
\gamma(r)(1-|\wt|^{-r})]^2+1}\right. \nonumber\\
&+&\left.\frac{5}{[\beta(r)-\gamma(r)(1-|\wt|^{-r})]^2+25}\right\}.
\eea The key point here is that equation (5.12) is independent of
$\Tt$, showing explicitly that for
$|\wt|\gg\mbox{max}(1,\Tt)$ the scaling spectrum indeed coincides
with its $T=0$ limit \cite{gl03}.

Second, equation (5.12) shows that the leading large $\wt$ behaviour
$(|\wt|^r\gg 1)$ of the GFL scaling spectrum is \bea \pi
\wm^rD(\w,T)&\sim&\frac{1}{2|\wt|^r}\left\{\frac{1}{[\beta(r)+
\gamma(r)]^2+1}\right. \nonumber\\
&\ \ \ +&\left.\frac{5}{[\beta(r)- \gamma(r)]^2+25}\right\} \eea
-- i.e.\ $\wm^rD(\w,T)\propto|\wt|^{-r}$ 
which, for small $r$, is very slowly varying.  The onset of this  powerlaw
behaviour for $|\w|^r\gg 1$ is readily seen in the inset to figure 3.

Finally, since $\gamma(r) =4/\mbox{sin}(\pi r) \sim 4/\pi r$ and
$\beta(r)=\mbox{tan}(\frac{\pi r}{2}) \sim \frac{\pi r}{2}$ for $r<<1$,
the leading low-$r$ behaviour of the high-frequency
asymptotic `tails' equation (5.13) is \be \pi \wm^rD(\w,T)\sim
\frac{3\pi^2r^2}{16}|\wt|^{-r}\ee precisely as found hitherto for
$T=0$ \cite{gl03}; and since the scale $\wm^r$ drops out of equation (5.14),
the low-$\w$ behaviour of the $T=0$ QCP spectrum itself (where $\wm=0$) follows 
immediately as \cite{gl03}
\be
\pi D(\w,0)=\frac{3\pi^2r^2}{16}|\w|^{-r}
\ee
which result is believed to be asymptotically exact as $r \ra 0$.

\subsection{LM phase}

  While the general form equation (5.10) for $T^{r}D(\w,T)$ naturally applies
to both phases, symmetry is not of course restored for $U>U_{c}$ in the
LM phase \cite{gl03,gl032}: the renormalized level $\st_{\up}^{\subr}(0,0)$
is non-zero.
But it necessarily
vanishes as $U\rightarrow U_{c}+$ and the LM$\rightarrow$GFL transition is
approached, and in consequence determines a low-energy scale $\wl$ characteristic of
the LM phase as discussed in \S 3.1. Given explicitly by 
\be \wl=
[|\st_{\up}^{\subr}(0,0)|/\gamma(r)]^{\frac{1}{r}}
\ee 
this is the natural scaling counterpart to $\wm$ in the GFL phase, such that the
$T=0$ LM phase spectrum scales universally as a function of
$\wt = \w/\wl$  \cite{gl03,gl032}.

  The required results for $\tilde{\Sigma}_{\uparrow}(\w,T)$ in the LM phase
are obtained most directly from their counterpart equations (5.8,9) in the GFL
phase (in which $1/\tilde{T} = \wm/T$ and $1/\wt =\wm/\w$),
by setting $\wm =0$ therein --- recall that
the spin-flip scale vanishes identically throughout the LM phase, where
$\chi^{+-}(\w,T)\simeq\mbox{Im}\Pi^{+-}(\w)=\pi\delta(\w)$. Equation (5.8) 
then yields
\be ( T^r)^{-1}\Sigma_{\up}^{\subi}(\w,T)=4|\w'|^rf(\w').\ee
$T^{-r}[\tilde{\Sigma}^{R}_{\up}(\w,T) -
\tilde{\Sigma}^{R}_{\up}(0,0)]$ is likewise obtained by setting $\wm =0$
in the right side of equation (5.9). Using equation (5.16) for
$\tilde{\Sigma}^{R}_{\up}(0,0) = - |\tilde{\Sigma}^{R}_{\up}(0,0)|$,
this gives
\be (T^r)^{-1}\st^{\subr}_{\up}(\w,T)=-\gamma(r)[\Tt^{-r}+
|\w'|^r]-\frac{4}{\pi}H(\w')\ee 
in which $\Tt=T/\wl$ (and $\w' = \w/T \equiv \wt/\Tt$).

\begin{figure}
\begin{center}

\psfrag{yaxis}[bc][bc]{${\wl}^rD(\w, T)$}
\psfrag{xaxis}[bc][bc]{$\wt$}
\psfrag{t1}[bl][bl]{$\tilde{T}=0.1$}
\psfrag{t2}[bl][bl]{$\tilde{T}=1$}
\psfrag{t3}[bl][bl]{$\tilde{T}=10$}
\psfrag{t4}[bl][bl]{$\tilde{T}=100$}
\psfrag{t5}[bl][bl]{$\tilde{T}=0$}

\epsfig{file=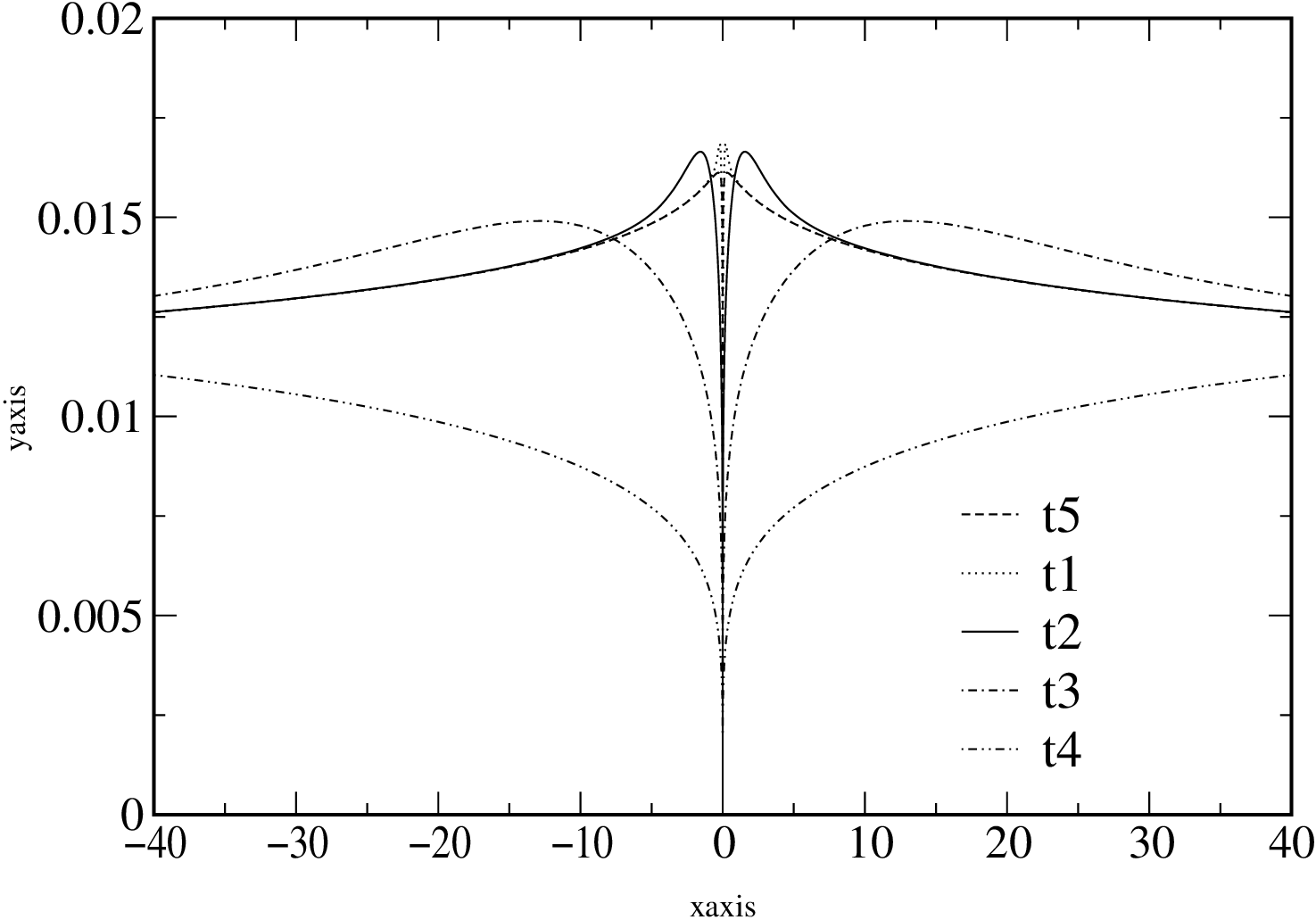,width=6cm, angle=270} \caption{LM phase scaling spectrum
$\wl^rD(\w, T)$ versus $\wt=\w/\wl$ for $r=0.2$ and a sequence of
temperatures $\Tt =T/\wl$.} \end{center}
\end{figure}

Equations (5.17,18) and (5.10) generate the LM phase scaling spectrum. It is
thus seen to be of form
\be
\pi T^rD(\w,T)=S_{\mbox{\ssz{LM}}}\left(\frac{\w}{T},\frac{1}{\Tt}\right) \ee
scaling universally in terms of $\w'$ and $\Tt$; or, entirely equivalently,
that $\wl^{r}D(\w,T)$ scales in terms of $\wt =\w/\wl$ and $\Tt$.
  The scaling spectra so obtained are shown in
figure 5 for a range of $\Tt$.  It is again readily shown that
for
$|\wt|\gg\Tt$ the LM phase scaling spectra are effectively
independent of $\Tt$ and coincide with the $\Tt=0$ limit \cite{gl03}, and 
from which the low-$\w$ behaviour of the $T=0$ QCP spectrum (equation (5.15))
correctly follows by taking $\wl\ra 0$.

\subsection{Quantum critical point} 

 We turn now to the quantum critical point itself. It has been shown in
\S 3  that the QCP spectrum must exhibit pure $\w' =\w/T$ scaling, \emph{ie} that
$\pi T^rD(\w,T)=S(\w')$. 
In \S s 5.1,2 we have also shown that this general 
behaviour is correctly recovered by the local moment approach (which is distinctly 
non-trivial, bearing in mind that
the QCP is an interacting, non-Fermi liquid fixed point).
Our aims now are twofold. First to obtain explicitly the scaling function $S(\w')$
arising within the LMA for $r\ll 1$;
and then to show that it is uniformly recovered from
GFL/LM scaling spectrum as $\Tt=T/\w_*\ra \infty$ (we remind the reader that
$\w_* \equiv \wm$ in the GFL phase and $\w_* \equiv \wl$ for the LM phase).

Taking $\Tt^{-1}=0$ in equations (5.17,18) (or equations (5.8,9))
gives 
\be 
(T^r)^{-1}\Sigma_{\up}^{\subi}(\w,T)=4|\w'|^rf(\w')\ee
\be ( T^r)^{-1}\st^{\subr}_{\up}(\w,T)=-\gamma(r)
|\w'|^r -\frac{4}{\pi}H(\w').\ee 
The QCP scaling spectrum $\pi T^r
D(\w,T)\equiv S(\w')$ now follows directly from equation (5.10).

\begin{figure}
\begin{center}
\psfrag{yaxis}[bc][bc]{$T^rD(\w,T)$}
\psfrag{xaxis}[bc][bc]{\ \ \ \ \ \ \ $\w/T$}
\psfrag{a}[bc][bc]{$|\w'|^{r}$}
\psfrag{b}[bc][bc]{\ \ \ \ \ \ \ \ $|\w'|^{-r}$}
\epsfig{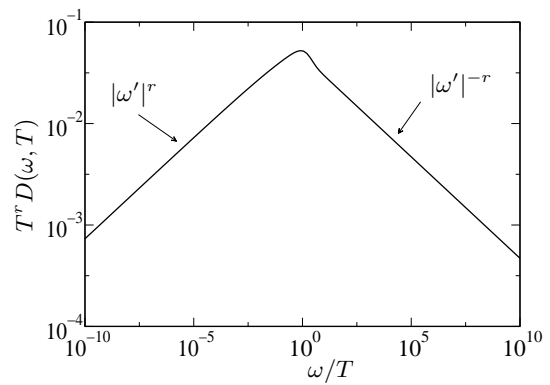} \caption{QCP scaling spectrum
$T^rD(\w,T)$ versus $\w' = \w/T$ for $r=0.2$.  $\w'\simeq 1$ marks a
crossover from $|\w'|^{-r}$ behaviour for $|\w'|\gg 1$ to $|\w'|^r$
behaviour for $|\w'|\ll 1$. } \end{center} \end{figure}

Figure 6 shows the resultant QCP spectrum for $r=0.2$, as a
function of $\w'=\w/T$ on a logarithmic scale.  From this it is clear
that two distinct power law behaviours dominate.
For $|\w'|\ll 1$, where $H(\w')\sim
r^{-1}(1-|\w'|^r)+c$ (with
$c=\mbox{ln}(\frac{\pi}{2}\mbox{exp}(-C))\simeq -0.125$ a constant,
and $C$ Euler's constant), it follows that \be \pi T^rD(\w,T)\stackrel{|\w'|\ll
1}{\sim}\frac{3\pi^2r^2}{16}|\w'|^r.\ee This behaviour is clearly seen
in the figure 6, and in practice sets in for $|\w'|\lesssim 1$.

For $|\w'|\gg 1$ by contrast, where $H(\w')\simeq 0$ may be neglected, equations
(5.10,20,21) give 
\be\pi T^rD(\w,T)\stackrel{|\w'|\gg
1}{\sim}\frac{3\pi^2r^2}{16}|\w'|^{-r}.\ee 
This form is also
clearly evident in figure 6, in practice setting in above 
$|\w'|\gtrsim 1$.  Such
behaviour is also precisely the general form deduced in section 3 (equation (3.9)) 
on the assumption that the limits $\w_*\ra 0$ and $T\ra 0$ commute;
with the explicit $r$-dependence $C(r) = 3\pi^{2}r^{2}/16$, which we believe
to be asymptotically exact as $r \ra 0$,  obtainable 
because the result now arises from a microscopic many-body theory.

\begin{figure*}
\begin{center}
\psfrag{xaxis}[bc][bc]{$\w/T$}
\psfrag{yaxis}[bc][bc]{$T^rD(\w, T)$}
\epsfig{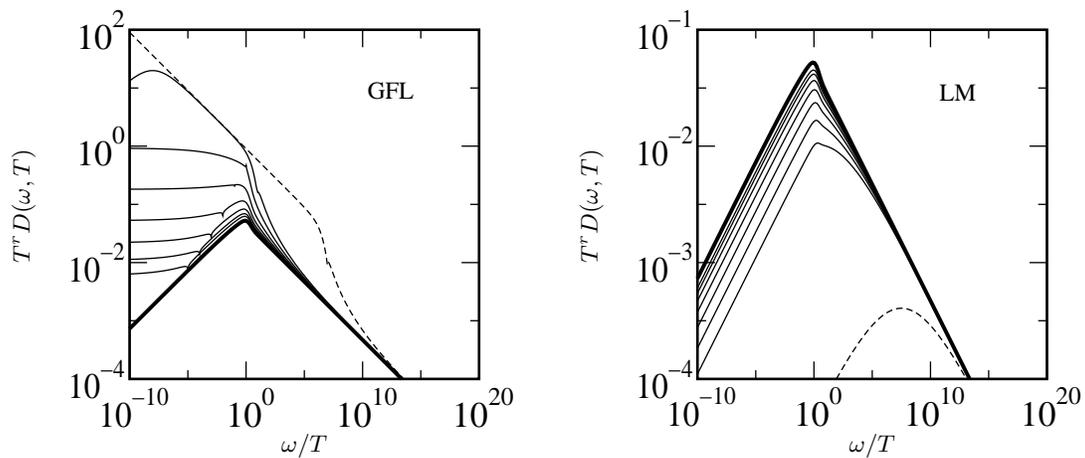} \caption{The left (right) panel shows
the GFL (LM) phase scaling spectrum
$T^rD(\w,T) \equiv S_{\mbox{\ssz{GFL}}}(\w', \tilde{T}^{-1})$ versus
$\w'=\w/T$ for  $\tilde{T} = 10^{-1}$,
$10^0$, $10^1$, $10^2$, $10^3$, $10^4$, $10^5$ and $\infty$
(in sequence top$\ra$bottom for the GFL phase and bottom$\ra$top
for LM). In
either case the QCP spectrum
$\tilde{T}=\infty$ is shown as a thick line; the dotted line shows the
$\tilde{T}\ra 0$ spectrum.} \end{center}
\end{figure*}

As discussed in section 3 and above, the GFL/LM scaling spectra may be
expressed as functions of $\w'=\w/T$ for any given $\Tt$.  One
obvious question then arises: how do the GFL/LM scaling spectra
evolve with temperature $\Tt$ to approach their ultimate limit of
the QCP spectrum as $\Tt\ra \infty$?  This is illustrated in
figure 7 where the analytic results of sections 5.1 and 5.2 are
replotted versus $\w'=\w/T$ for a sequence of increasing
temperatures $\Tt=T/\w_{*}$.  The QCP scaling spectrum is shown as a thick dashed
line.  From figure 7 we see that --- for all $\Tt$ --- the GFL and LM scaling
spectra coincide both with each other and with the QCP scaling
spectrum for $|\w'|\gg 1$.  This is a reflection of two facts.
First, as discussed in sections 5.1 and 5.2, that the high-frequency behaviour
of the GFL and LM phase spectra coincide with their $\Tt=0$
limits.  And second, that the high-frequency tails of the $T=0$
GFL/LM scaling spectrum coincide with those of the $T=0$ QCP
spectrum.

Most importantly, however, figure 7 shows that the QCP scaling
spectrum is obtained `uniformly' from the GFL/LM phase scaling
spectrum upon increasing temperature $\Tt$.  That is, 
with progressively increasing $\Tt$ the
GFL/LM phase scaling spectra ever increasingly coincide with the
QCP scaling spectrum (thick line) over a larger and larger
 frequency interval,
such that in the limit $\Tt\ra\infty$ the QCP spectrum for all
$\w/T$ is obtained smoothly.  This is an important result,
providing as it does a direct connection between the QCP scaling
spectrum and scaling spectra in the GFL or LM phases at any finite
$\Tt$.

\section{Summary} We have considered a local moment approach to the
pseudogap Anderson impurity model, close to the symmetric quantum critical
point where the Kondo resonance has just collapsed.  Building on previous
work \cite{gl03,gl032,ld02}, we have focused on
single-particle dynamics at finite temperature, obtaining an analytical
description of the finite-$T$ scaling behaviour for small $r$, in both
generalized
Fermi liquid (Kondo screened) and local moment phases. 
A key result obtained on general grounds is that pure $\w/T$-scaling
obtains at the QCP itself, consistent with an interacting fixed point and  recent
results for the local dynamical susceptibility \cite{ingsi02}.  
We have succeeded both in obtaining explicitly the QCP scaling spectrum, and
in understanding its continuous emergence with increasing $T/\w_*$ from the
scaling dynamics appropriate to the Kondo screened and local moment phases
on either side of the quantum phase transition.

Related results have also been obtained for the PAIM with finite local
magnetic field, $h=g\mu_{\mbox{\ssz{B}}}H_{\mbox{\ssz{loc}}}$ \cite{ld01a,ldb01}; i.e.\
pure
$\w/h$-scaling of the single-particle spectrum at the QCP.  These 
will be discussed elsewhere. 

\hspace{1cm}

\noindent{\it Acknowledgements} \newline
The authors would like to express their
appreciation to the EPSRC, Leverhulme Trust and Balliol College Oxford
for support; and to Kevin Ingersent for helpful discussions regarding the
present work.

%
%
%
%

\end{document}